\documentclass{iaus}
\usepackage{graphicx}
\def\ltsima{$\; \buildrel < \over \sim \;$}
\def\simlt{\lower.5ex\hbox{\ltsima}}
\def\gtsima{$\; \buildrel > \over \sim \;$}
\def\simgt{\lower.5ex\hbox{\gtsima}}

\title[The Thin and Thick Disks] 
{Kinematical \& Chemical Characteristics of the Thin and Thick Disks}

\author[Rosemary F.G.~Wyse]   
{Rosemary F.G.~Wyse}

\affiliation{Department of Physics \& Astronomy, Johns Hopkins University, \\ Baltimore, MD 21218, USA  \\ email: {\tt wyse@pha.jhu.edu}}

\pubyear{2008}
\volume{254}  
\pagerange{xxx}
\jname{The Galaxy Disk in Cosmological Context}
\editors{J.~Andersen, J.~Bland-Hawthorn \& B.~Nordström, eds.}
\begin{document}

\maketitle

\begin{abstract}
I discuss how the chemical abundance distributions, kinematics and age
distributions of stars in the thin and thick disks of the Galaxy can
be used to decipher the merger history of the Milky Way, a typical
large galaxy. The observational evidence points to a rather quiescent
past merging history, unusual in the context of the `consensus'
cold-dark-matter cosmology favoured from observations of structure on  scales larger than individual galaxies. 


\end{abstract}

\firstsection 
\section{Introduction: Formation of Thin and Thick Disks in {$\Lambda$}CDM}
 Dissipational collapse of a gas-rich system is an important ingredient
in establishing the thin disks so prevalent today.  In the context of
hierarchical clustering scenarios, such as {$\Lambda$}CDM, gaseous
mergers are required to produce disks (Zurek, Quinn \& Salmon 1988;
Robertson et al.~2006).  Centrifugally supported, extended disks also
are only produced if angular momentum is largely conserved during
collapse within the dark halo (Fall \& Efstathiou 1980; Mo, Mao \&
White 1998).  However, angular-momentum transport and evolution, for
example due to gravitational torques and tidal effects, are natural
during the mergers inherent in {$\Lambda$}CDM, leading to low angular
momentum, very concentrated disks (Navarro, Frenk \& White 1995).  The
typical merger history (of dark haloes) is fixed by the dark matter
power spectrum, so with {$\Lambda$}CDM additional baryonic processes
are introduced to implement `feedback', to both suppress early star
formation and prevent dissipation, delaying disk formation until after
the epoch of most active merging (cf.~Simon White's talk).  Later
mergers into the disk will heat the thin disk, with minor mergers
producing a thick stellar disk (some thin stellar disk component
persists, e.g. Kazantzidis et al.~2007) and also driving gas into the
central regions to build up a bulge (Mihos \& Hernquist 1996).

During a merger, orbital energy is absorbed into the internal degrees
of freedom of the merging systems, and orbital angular momentum is
redistributed, some absorbed by the larger system, and some being lost
to the system. The evolution of a satellite, and of its orbit, as it
merges with, and is assimilated by, a larger system depends on the
relative masses (the dynamical friction timescale on which the
satellite sinks to the center scales like the mass ratio, being
shorter for more massive satellites), on the relative density profiles
of the satellite and host (denser satellites can survive tidal effects, leading to mass loss and disruption, closer to the
center of the larger system), and on the initial orbital parameters
(e.g. sense of rotation, inclination angle to the plane of the host,
periGalacticon and orbital eccentricity).  The effect of the
satellite(s) on a pre-existing stellar disk is also a sensitive
function of the satellite's properties and orbit.  The mechanism by
which a thin disk is heated by the merging process is a combination of
local deposition of orbital energy from the satellite(s), local scattering (in which azimuthal
streaming motions are transformed into random motions), and resonant
excitation of modes in the disk, providing heating on a more global scale (e.g.~Quinn \& Goodman 1986; T\'oth \& Ostriker 1992; Sellwood, Nelson \& Tremaine 1998).

Early simulations focussed on the impact of one minor merger
(e.g.~Quinn, Hernquist \& Fullagar 1993; Velazquez \& White
1999). These produced a plausible thick disk, similar in structure to
that of the Milky Way (Gilmore \& Reid 1982; see also Gilmore \& Wyse
1985 for kinematics and an order-of-magnitude estimation of the mass
of satellite needed), from a merger of a robust (dense) satellite with
a mass ratio to the stellar {\it disk\/} (not to the total mass) of
10--20\%, for a range of initial orbital parameters.  Simulations of
the merging of cosmologically motivated ensembles of satellites are
more relevant, and also find significant heating of a pre-existing
stellar disk, over an extended period of time (e.g.~Hayashi \& Chiba
2006; Kazantzidis et al.~2007, also this volume).  The satellites in
these later simulations have a orbital distribution similar to that of
subhaloes identified in dissipationless $\Lambda$CDM cosmological
simulations, with a mean ratio of initial apocenter to pericenter
distance of around 6:1 (e.g. Ghigna et al.~2000; Diemand, K\"uhlen \&
Madau 2007). The distribution of pericenter distances is also
important, since satellites with pericenters that are significantly
beyond the disk (larger than say 10 disk scale lengths, see Fig.~2 of
Hayashi \& Chiba 2006) do not couple effectively to the disk. In
addition to realistic orbits, the internal density profiles of the
satellites are critical, since fluffy satellites are disrupted early
and provide little heating (Huang \& Carlberg 1997).  High-resolution,
N-body simulations of the formation of the dark halo of a `Milky Way
galaxy' with the CDM power spectrum have demonstrated that a
significant population of substructure is indeed dense enough to
persist and survive many pericenter passages.  The shapes of the mass-
and velocity-functions of subsystems are reasonably independent of
redshift and at $z=0$ are well-established as power laws (see
convergence tests in Reed et al.~2005; S.~White's talk at this
conference). The amplitudes depend on resolution, with the number of
satellite dark haloes still increasing with increased resolution
(S.~White, these proceedings); `overmerging', particularly in the
central regions of the larger host galaxy halo, can artificially
reduce substructure.  Indeed, the population of subhaloes within the
analogue of the solar neighborhood is not yet well-established, even
in pure dark-matter simulations (the addition of baryons will increase
central densities). However, the present generation of (baryon-free)
simulations imply that robust satellites penetrate the region of the
disk (e.g.~Diemand, K\"uhlen \& Madau 2007).  Simulations which model
gas physics and star formation also find that thick (and thin) disks
are produced. For example, the bulk of the younger stars (ages less
than 8~Gyr) in the thick disk in the simulation of Abadi et al.~(2003)
are produced by heating of the pre-existing stellar disk (we return to
the older stars in section~2 below).

The most massive satellite provides the greatest heating, with the
scaling such that the increase in scaleheight (or, equivalently, in the
square of the vertical velocity dispersion) is proportional to the
square of the mass of the satellite (Hayashi \& Chiba 2006). For an
ensemble of satellites with the differential mass function seen in CDM
simulations, namely a power-law with slope close to $-2$
(e.g.~Diemand, K\"uhlen \& Madau 2007), the cumulative 
heating is also dominated by the
most massive systems (see also White 2000).
The substructure distribution at earlier times contains more
satellites of higher mass, since these are more affected by dynamical
friction, which brings them into the central regions where they are
more effectively stripped of mass from their outer parts (e.g.~Zentner
et al.~2005).  These massive
satellites, after this shriking of their orbits, can be more effective
at heating the thin disk, prior to their demise.  Thus at early times
the satellite distribution is more concentrated than is the host dark
matter halo, while at the present day it has evolved to be less
concentrated.  One must allow for this evolution of the mass function
and orbital parameters, rather than simply adopting a surviving
satellite retinue from the end-point of a simulation, which minimizes
the predicted overall heating (as found by e.g.~Font et al.~2001).  Indeed
following the full merging history is preferable. Such analyses  imply that late
(after redshifts of unity) heating of thin stellar disks seems to be
inevitable in $\Lambda$CDM (e.g.~Abadi et al.~2003; Stewart et al.~2008; Kazantzidis,
this volume).  The current models (Stewart et al.~2008) show that over
the last 10Gyr, fully 95\% of galaxy haloes of present total mass
$10^{12}$~M$_\odot$ have accreted a system of mass equal to that of
the present-disk ($5 \times 10^{10}$M$_\odot$ -- their simulation resolution limit is $10^{10}$~M$_\odot$); the mass ratio of the
substructure to that of the stellar disk at the epoch of accretion is
the more important ratio for disk heating 
and such a large mass is highly likely to cause severe heating. More numerous, lower-mass mergers are also expected, continuing to late epochs.

Gas physics can also play a role in the formation of the thick disk,
in terms of slow settling to the disk plane (Burkert, Truran \&
Hensler 1992) or a starburst in a rapidly changing potential such as a
gas-rich merger (Robertson et al.~2006; Brook et al.~2007). The 
latter mechanism has some observational support at high redshift (see
Elmegreen's and Genzler's contributions to this volume).  One must of course also take account of adiabatic compression and heating of an existing 
thick disk by subsequent slow accretion of gas to buildup the thin disk (cf.~T\'oth \& Ostriker 1992; Elmegreen \& Elmegreen 2006). Here I will
focus on the -- apparently inevitable -- late heating of thin stellar
disks, as a probe of $\Lambda$CDM.  The important issues are the
predicted chemical abundance and age distributions of stars in the thin and thick disks,
given a typical merger history, and how they compare with the
observations. 

\section{Evidence for Minor Mergers in the Past}

Satellites that are accreted are in general only partially
assimilated, with `shredded satellite' debris maintaining some
coordinate-space coherence for a few orbits, kinematic coherence for
longer, and with persistent stellar population signatures, in terms of
their chemical abundances and stellar age distributions, allowing identification over a Hubble time.  In general
one can expect satellite debris to be deposited along the (evolving)
orbit of its center-of-mass, leading to a prediction that
former-satellite member stars will populate the thick disk -- halo
interface at a range of Galactocentric radii (see e.g.~Fig.~19 of
Huang \& Carlberg 1997, Fig.~9 of Abadi et al.~2003, Fig.~3 of Meza et al.~2005), again with
details depending on the satellite mass, internal density distribution
and initial orbit.  Indeed, it has been predicted that a large
fraction of the old stars in the thick and thin disks consists not of
stars formed {\it in situ\/} but rather stars accreted from satellites
many Gyr after they were formed (Abadi et al.~2003), with debris from each
satellite populating a different radial range, and the parent satellite having been brought to a circular orbit prior to mass loss. This late
(redshifts $z < 1$) accretion of old stars, directly into the disks, on high angular momentum orbits, would allow reconciliation of the 
delayed formation of disks, required in $\Lambda$CDM as discussed
above, with the presence of old `disk' stars.

Only high mass, dense (robust) satellites can experience efficient
circularization of their initial orbits through dynamical friction.
Accretion of such objects into the plane of the disk should also cause
heating of the thin disk that has formed by the epoch of their
accretion, leading to a thick disk component with a stellar age
distribution that reflects the thin disk star formation history up to
that epoch. The derived star-formation history of the (local) thin
disk is fairly smooth and continuous from the earliest times,
corresponding to the lookback time of $\sim 10-12$~Gyr that equals the
ages of the oldest stars in the thin disk (Binney, Dehnen \& Bertelli
2000). Taking Abadi et al.~as an example, with a significant accretion
event at $z = 0.73$, i.e.~a look-back time of $\sim 7$~Gyr, one
expects stars in the thick disk as young as 7~Gyr, rather than a
uniformly old population.  This is indeed what they find in that simulation (their Fig.~8). 

Note that the Monoceros Stream
(Newberg et al.~2002) and Canis Major overdensity have been
interpreted in terms of the in-plane accretion of a dwarf galaxy
(Pe\~narrubia et al.~2005). The null detection of an associated
over-density of RR Lyrae stars by Mateu et al.~(poster this conference)
would be unexpected in this scenario, but is consistent with dynamical
instabilities -- warp, flare, spiral arms -- in the outer
disk. \footnote{`The Monoceros Stream' was first identified by Corlin
(1920; his Table~2) as a local moving group. While clearly a different
feature, this first Monoceros Stream may hold lessons for the
interpretation of the current Stream.} 

\subsection{The Age Distribution of Stars in the Thick Disk}

All available observations are consistent with a dominant old age for
the stars of the thick disk of the Milky Way, where `old' means as old
as the globular clusters of the same metallicity (e.g.~47~Tuc), or at
least 10~Gyr, and probably 12~Gyr. The most reliable evidence comes
from looking at the turnoff for {\it in situ\/} thick disk stars,
several thin disk scale heights above the plane (to minimize
contamination by outlier thin disk stars), as a function of
metallicity.  Samples analysed in this way 
show very few stars bluer than
the 10-12~Gyr turn-off colour at a given metallicity, for both thick
disk and stellar halo (Gilmore, Wyse \& Jones 1995 and
references therein).

\begin{figure}[h]
\begin{center}
 \includegraphics[angle=270,width=3.6in]{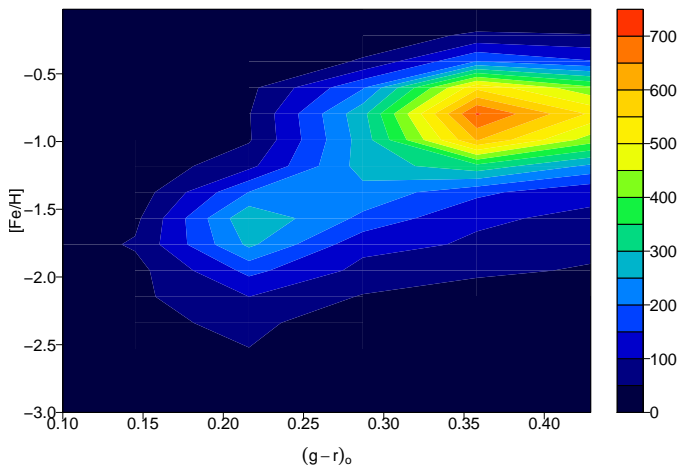} 
 \caption{Contour plot of the locations of 8,600 faint F/G stars in
 the plane of de-reddened colour and metallicity. The rather abrupt
 turn-offs of each of the thick disk and halo are apparent, with few
 younger stars, which, if present, would occupy the upper left portion
 of the plane.} \label{fig1}
\end{center}
\end{figure}

The data for 8,600 faint F/G dwarfs with metallicities from 
intermediate-resolution spectroscopy (obtained with the AAT/AA$\Omega$
multi-object spectrograph) are shown in Fig.~1 (Wyse, Gilmore
\& Norris, in preparation), where the dominant turn-off is seen.   The turn-off colour of the thick disk (and the
bluer turn-off colour of the more metal-poor stellar halo) from the
full Sloan Digital Sky survey imaging data, covering a significant
fraction of the sky, also implies this old age for both thick disk
and stellar halo, more globally across the Galaxy (e.g.~Ivezic et
al.~2008).  Local samples -- making use of Str\"omgren photometry --
are more susceptible to contamination by thin disk stars with extreme
kinematics, such as due to three-body interactions, and require
careful analysis to isolate a clean thick disk sub-sample. With this
caveat, such samples are in general consistent with this dominant old
age (e.g.~Str\"omgren 1987; Nordstr\"om et al.~2004; Reddy et
al.~2006; Schuster et al.~2006).

Such a high value for the age of the dominant stellar population of
the thick disk has major implications for the minor merging history of
the Milky Way.  As noted above, stars of all ages are found in the
local thin disk, with a derived star formation history that extends
back to earliest times. Thus, if the thick disk originated through
merger-induced heating of a pre-existing thin stellar disk, the last
significant (defined as having a mass ratio to the disk of $\sim
20\%$, and surviving to interact with the disk) dissipationless merger
can be dated by the age distribution of stars in thick disk: if the
typical thick disk star is old, then the last such merger was long
ago, with an age greater than 10~Gyr setting a limit of  no significant merger
activity and heating since a redshift of $ \simgt 2$.  This scenario
also requires there to have been an extended thin stellar disk in
place at $z \sim 2$ (even allowing for some radial mixing
subsequently).

Mergers do not only heat thin disks, but can also drive gas and stars
into the central regions to build-up the bulge. It may then be no
coincidence that the old age of the dominant stellar population in the
bulge of the Milky Way, again 10--12~Gyr (e.g.~Zoccali et al.~2003;
Feltzing \& Gilmore 2001), provides a consistent limit on merger
activity. As noted previously (Wyse 2001), it could be that the merger
that created the thick disk induced gas inflow and an associated
starburst to form the (bar/)bulge {\it in situ\/}.

\subsection{What fractions of the disks can be direct stellar accretion from satellites?}

\subsubsection{Kinematic Constraints}
Interloper stars, accreted during the merger with a
satellite galaxy, should be identifiable, probably with different
kinematics, chemical abundances, age and spatial distributions from
stars formed {\it in situ.} It might be remembered that the
Sagittarius dwarf spheroidal was discovered serendipitously during a
spectroscopic survey of the Milky Way bulge (Ibata, Gilmore \& Irwin
1994) by the distinct kinematics and colour distributions of its member
stars.  In the scenario whereby the thick disk results from heating of
a pre-existing thin disk by minor mergers, the `shredded-satellite'
stars should be distinguishable from `heated thin-disk stars'.  

Such
satellite debris was identified with  `thick disk' stars, observed 
 several kiloparsecs above the thin disk
plane, in two lines-of-sight at longitude $\sim 270^\circ$, on orbits of  significantly lower angular
momentum than the standard thick disk star -- with a lag
in mean azimuthal streaming of $\sim 100$~km/s behind the Sun,
compared to the standard thick disk lag of $\sim 40$~km/s (Gilmore,
Wyse \& Norris 2002).  A similar population was found  among the Galactic field stars  along lines of sight to
dwarf spheroidal galaxies, at widely separated
lines-of-sight (Wyse et al.~2006).  The radial velocity histogram from
the subset of our AAT/AA$\Omega$ data at $\ell \sim 270^\circ$, where the
line-of-sight velocity has a significant contribution from the
azimuthal streaming,  is 
shown in Fig.~2 (Wyse, Gilmore \& Norris, in preparation).
There is clearly again a significant population with a rotational lag of
$\sim 100$~km/s.  These stars have a broad range of metallicity
(derived from the Ca II K line), down to $-3$~dex.  A full analysis is 
underway.

\begin{figure}
\begin{center}
 \includegraphics[angle=270,width=3.5in]{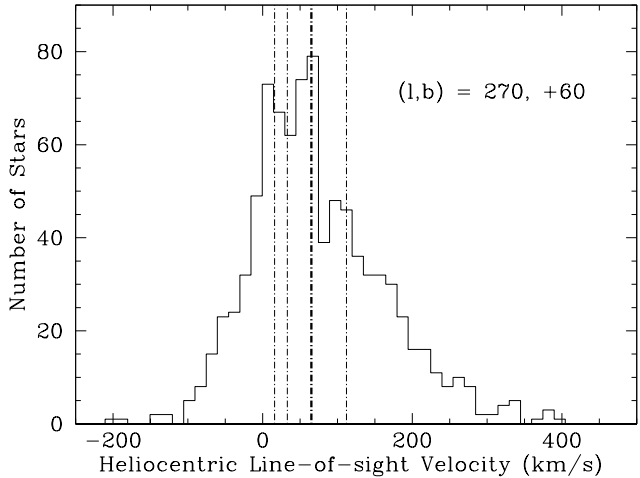} 
 \caption{Line-of-sight velocity histogram for the $\sim 900$ faint F/G stars at $\ell = 270^\circ$ from our wide-area spectroscopic survey with AA$\Omega$ (some 12,000 stars in total).  The predicted mean velocities for the standard (old) thin disk, thick disk and stellar halo are shown as fainter vertical dot-dashed lines (left to right), while the heavier vertical dot-dash line is the predicted mean for a component lagging behind the Sun by 100~km/s.  This clearly matches the peak velocity.} 
   \label{fig2}
\end{center}
\end{figure}

While we interpreted our results in terms of
discontinuous kinematics distinguishing `shredded-satellite stars'
from `heated thin-disk stars' (true thick disk in this picture), others
with similar quality spectroscopic data have modelled their velocities by
smooth gradients as a function of height from the disk plane
(e.g.~Chiba \& Beers (2000) with a similar sample size of around one
thousand stars).  The Sloan Digital Sky Survey photometric data for $\sim 60,000$~F/G stars at the
NGP ($b > 80^\circ$) have been analysed together with proper motions
by Ivezic et al.~(2008), using photometric metallicity determinations
to derive distances and hence tangential velocities (decomposable into
velocity towards the Galactic center and azimuthal streaming
velocity). They deduce a steep gradient in rotational lag with Z-height for the thick disk, similar in amplitude to that of Chiba \& Beers (2000), namely $\sim 30$~km~s$^{-1}$/kpc. It will be interesting to try alternative models to describe the data, while noting that the heated thin-disk stars in the minor-merger simulations of Hayashi \& Chiba (2006) shows a vertical gradient in rotational velocity of comparable amplitude.

\subsubsection{Age Constraints}

The surviving satellites in the Local Group all contain old stars,
consistent with star formation in all small galaxies being initiated around
the epoch of reionization (e.g.~Hernandez, Gilmore \& Valls-Gabaud
2000; Dolphin 2002).  Most of the luminous satellites had an extended
and fairly continuous star-formation history, albeit non-monotonic,
and contain a dominant intermediate-age population, contrasting with
the dominant old ages seen in the bulge, thick disk and stellar halo
of the Milky Way. While satellites accreted early will therefore
contain stars with the same age distribution of the non-thin-disk
components of the Milky Way, satellites accreted later may be expected
to contain significantly younger stars. Accretion of typical luminous
satellites to form more than a few percent of the stellar halo is then limited to epochs prior to a look-back time of
$\sim 10$~Gyr, or again a redshift of $\sim 2$ (Unavane, Wyse \&
Gilmore 1996). The similar old age of the thick disk stars produces similar constraints.   Systems that contain uniformly old stellar
populations, such as the Ursa Minor dSph, could of
course be assimilated into the Galaxy at any epoch and would not be
distinguishable by the stellar age distribution (or their stellar mass function; Wyse et al.~2002), but only a small
fraction of the stars in dwarf galaxies now are as old as the stars in
the Ursa Minor dSph, and the stellar mass of the Ursa Minor dSph, $\sim 10^6$~M$_\odot$, is  a tiny fraction of even the stellar halo.  

\subsubsection{Overall Metallicity Constraints}
The metallicity distribution of the local thick disk is distinct from
any of the other stellar components, but of course the tails overlap
(see e.g. Wyse \& Gilmore 1995). The mean metallicity of the local
thick disk, expressed as an iron abundance, is around one-quarter of
the Solar value. The luminosity-metallicity relation for galaxies in
the Local Group (e.g.~Mateo 1998) implies that only the most luminous
satellites can self-enrich to this value, suggesting that the
thick-disk stars formed in a system of relatively deep potential
well. The Large Magellanic Cloud has enriched to this level, and the
metallicity distribution of the inner disk of the LMC (Cole, Smecker-Hane \& Gallagher 2000) is similar to that of the local thick disk.  Further, the
total stellar masses are comparable.  However, the LMC had a much
slower enrichment history than did the (progenitor of) the thick disk,
and reached [Fe/H] $\sim -0.6$~dex only $\sim 5$~Gyr ago (Hill et
al.~2000).  The rapid enrichment of the thick disk points to a high early star formation rate, and chemical evolution in a system significantly
more massive than the LMC, suggestive that the overall Milky Way
potential was already established at $z \sim 2$, and the bulk of the star formation was {\it in situ}.

\subsubsection{Elemental Abundance Constraints}

Stars of different masses synthesize and eject different elements, on
different timescales, so that elemental abundances contain much more
information than does overall metallicity.  The latter is an integral
over past star formation and chemical enrichment, while the former
reflects the ratio of recent star formation rate (through enrichment
by Type~II supernovae, which evolve on timescales of $\sim 10^7$~yr
after birth of the progenitor star, faster than the typical duration of star formation) to past star formation
(e.g.~through Type~Ia supernovae, which evolve on timescales of
several times 10$^8$~years, up to a Hubble time, after the birth of the
progenitor stars). Massive stars, ending their lifes as core-collapse
(Type~II) supernovae, create and eject intermediate-mass elements, in
particular those synthesized by the addition of successive helium
nuclei, and known collectively as the `alpha-elements'.  The $r-$process
elements are also created in the high neutron-flux environments of
Type~II supernovae. Stars that are formed early in a star-forming
event, prior to significant Type~Ia supernovae activity, (not
necessarily early in absolute terms) will be enriched by only Type~II
supernovae. Provided there is good mixing of ejecta, and a massive
enough star-formation event for the massive-star Initial Mass Function
(IMF) to be fully sampled, the interstellar medium from which these
early stars form will be enriched with a characteristic ratio of
$\alpha$-elements to iron.  This characteristic ratio is set by the
massive-star IMF, since the mass of iron that is produced is
essentially independent of progenitor mass, while the mass of
$\alpha$-elements produced is a steeply increasing function of
progenitor mass, independent of progenitor metallicity (see e.g.~Fig.~1 in Gibson 1998; Kobayashi et
al.~2006).  Thus if the massive-star IMF were biased towards more
massive stars (remembering the relevant range is $\sim 8$~M$_\odot$ to
$\sim 100$~M$_\odot$), stars enriched by the resulting Type~II
supernovae only would show a higher value of [$\alpha$/Fe]. As
discussed in Wyse \& Gilmore (1992), IMF slopes that have been
proposed (for various reasons) predict values of this `Type~II plateau'
that can differ by greater than 0.3~dex, certainly an observable effect. Such differences  have not been observed, providing strong evidence against a variable IMF.  

\begin{figure}
\begin{center}
 \includegraphics[angle=270,width=3.5in]{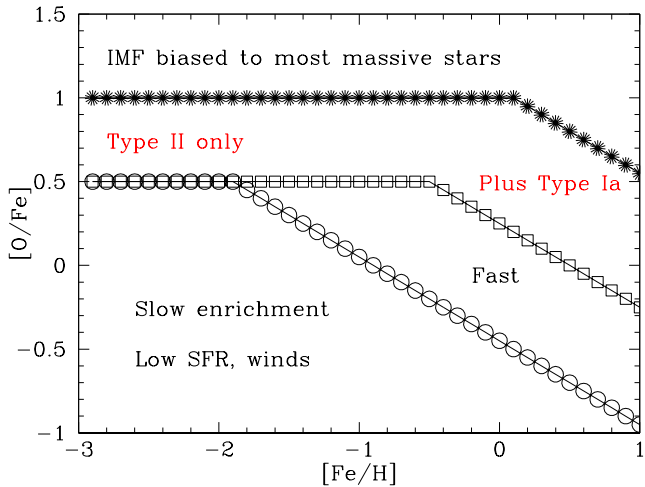} 
 \caption{Schematic elemental ratio pattern for self-enriching
 star-forming regions.   Stars formed at early times in the star-forming event
 are pre-enriched by only Type~II supernovae and with good mixing of
 ejecta into the interstellar medium, this pre-enrichment has a fixed
 value of [O/Fe]. The asterisks represent an IMF biased towards
 the most massive stars, while the open squares and circles represent a
 normal IMF. The values of the `Type~II plateau' reflect the differences in massive-star IMF. 
 } \label{fig3}
\end{center}
\end{figure}

Type~Ia supernovae are produced by accretion onto a massive white
dwarf in a binary system and each produce a relatively large mass of
iron, and a small mass in $\alpha$-elements. The signature of the
incorporation of the ejecta from Type~Ia in the element ratios of
long-lived low mass stars is a lower value of [$\alpha$/Fe] than the
Type~II plateau. Irrespective of the details of the model, the minimum
delay time after star formation, for a Type~Ia supernova, is given by
the time taken for an $\sim 8$~M$_\odot$ star (the most massive
progenitor in this case) to become a white dwarf, accrete sufficient
material to exceed the Chandrasekhar mass, and then explode. This is
the origin of the several times 10$^8$~yr delay noted above; the
shortest timescale for incorporation of significant iron into the ISM
and the next generation of stars is usually estimated as $\sim
10^9$~yr.  Lower-mass progenitors take longer to end up as white
dwarfs, and different binary systems can range tremendously in their
evolution and accretion times, leading to a long tail in delay times
(see e.g.~Matteucci \& Greggio 1986; Smecker \& Wyse 1991). The
enrichment by Type~Ia supernovae is set by a delay {\it time,} and the
iron abundance corresponding to that time depends on the efficiency of
chemical enrichment.  The rate of chemical enrichment depends on the
star formation rate, gas flows, and on the ability of the system to
retain metals. In the absence of a mechanism to remove individual
elements preferentially in a wind, none of these will modify the value
of the Type~II plateau, but will modify the iron abundance at which
the downturn from this plateau appears.

The situation is illustrated schematically in Fig.~2 (modified from
Wyse \& Gilmore 1993).   With a fixed  IMF, the value of [$\alpha$/Fe]
is fixed, for stars that form early, and for a normal IMF that value is $\sim +0.3$. Thus one expects the metal-poor stars in any self-enriching system 
to show such values.  Of course, if there are
subsequent bursts of star formation, so that Type~II supernovae 
dominate again, newly forming stars in that burst will also show
these enhanced values of [$\alpha$/Fe] (e.g.~Gilmore \& Wyse 1991 for models; Koch et
al.~2008 for application to the Carina dSph). 

It is clear that elemental abundance patterns reflect the IMF and star
formation histories of the star-forming systems. In particular,
systems like the dwarf spheroidal galaxies, with inefficient enrichment
over extended periods, should  show low values of [$\alpha$/Fe] at low
values of iron (Unavane, Wyse \& Gilmore 1996), as observed (Venn et
al.~2004). Each (surviving) satellite galaxy in the Local Group has
its own star-formation and enrichment history, leading to the
expectations of a unique pattern in elemental abundances for each
system.  The realisation of this expectation is demonstrated by
Geisler et al.~(2007; their Fig.~12), where the different loci in the
[$\alpha$/Fe]--[Fe/H] plane of Galactic stellar populations and of each dwarf satellite is
rather dramatic. The vast majority of stars in dwarf galaxies now all lie below the Galactic populations in this plane, with $0.2 > {\rm {[\alpha/Fe]}} > -0.3$.  The distinct patterns of stars in different satellites  means that one should be able to identify
candidate `interloper' stars from a given system, from their joint kinematics and elemental
abundance distributions, provided their parent system formed stars for
longer than $\sim 10^9$~yr, and the stars were accreted subsequently (see Nissen
\& Schuster, this volume, for an interesting discussion of `low-alpha'
stars with halo kinematics). 

\begin{figure}
\begin{center}
 \includegraphics[angle=270,width=3.5in]{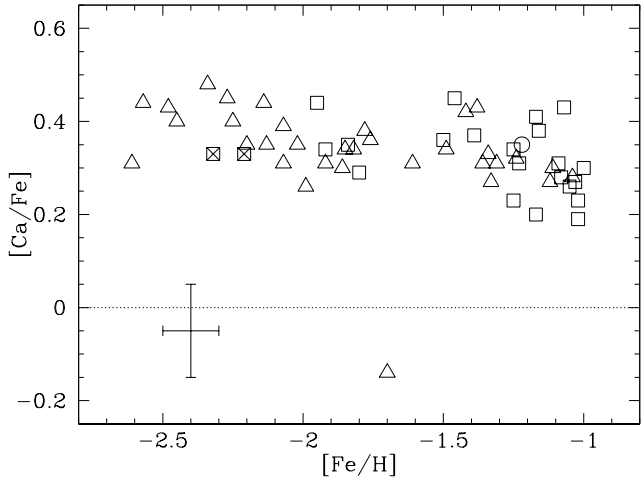}
 \caption{Elemental abundance ratios for the $\alpha$-element Calcium, in metal-poor stars ([Fe/H] $< -1$~dex)  selected from the RAVE catalogue on the basis of disk-like kinematics. The stars are assigned to populations using
 several criteria, based on the refined stellar parameters from
 high-resolution spectra. Circles are thin-disk stars, 
 squares are thick-disk stars and  triangles  are halo
 stars. The crossed- square symbols at [Fe/H] $< -2$~dex represent two 
 stars with uncertain thick disk--halo designation at present. The
 thick and thin disks clearly extend to low iron abundances, and those
 stars have enhanced [$\alpha$/Fe], unlike the bulk of stars in
 satellite galaxies, which have values of [$\alpha$/Fe] $<  0.2$ at these metallicities. We expect to double this sample this observing season.} \label{fig4}
\end{center}
\end{figure}

Thus late accretion of satellite galaxies into the plane of the disk
may be expected to produce disk stars with low iron abundance and low
[$\alpha$/Fe]. The RAVE spectroscopic survey of apparently bright
stars (Steinmetz et al.~2006, also this volume) provides an ideal
sample in which to look for candidates: stars with low metallicity but
disk-like kinematics. We (Ruchti et al., in preparation) have
initiated a programme to obtain follow-up high resolution spectroscopy
of such candidates, using the AAT, Magellan and the ESO 2.2m
telescope. These spectra provide improved stellar parameters and
elemental abundances. Reflecting the magnitude-limited nature of the
RAVE selection function, most of the candidates are giants, with
distances in the range of 500~pc to a few kpc.  Several criteria, with
different dependences on the distance estimates and kinematics (radial
velocities and proper motions), are used to assign each star to a
given population. This assignment is probabilistic in nature and so cannot be definitive for any one star; large samples are needed and we have been awarded further observing time this semester to obtain a statistically significant sample.  Elemental abundances are derived using the
methodology of Fulbright (2000) and  our results thus far for metal-poor stars are illustrated in Fig.~4. We
find that
the thick disk extends to at least [Fe/H] $= -2$~dex, and the thin disk
below $-1$~dex.  These metal-poor disk stars do not have the low
ratios of [$\alpha$/Fe] of the bulk of the population in dwarf
galaxies.  This limits the accretion of stars from satellite galaxies,
into the disks, to have occurred only very early, and is not
obviously consistent with late accretion, at epochs $z < 1$, as has been
proposed (Abadi et al.~2003). 

These data also serve to illustrate another unexpected and important
point: the small scatter about the `Type~II plateau', even at the
lowest abundances where only a handful of supernovae suffice to
provide the enrichment (see also Cohen et al.~2004; Cayrel et
al.~2004; Spite, this volume).  The different star-forming regions
that were the basis for the major stellar components of the Milky Way
were apparently enriched by stars with a fixed (massive-star) IMF, and
were well enough mixed, and massive enough, to average out the
different yields from supernovae of different progenitor masses.  The 
separation between different  components, and lack of scatter, within any one component (e.g.~Reddy, Lambert \& Allende-Prieto 2006; Bensby et al.~2007a),  
around the downturn signalling the contributions of Type~Ia supernovae, at least in samples probing a few kiloparsec of the solar circle, is hard to understand if many distinct small subsystems  contribute. 

\subsection{Substructure and Mergers into the Thin Disk}

As noted, it has been suggested that direct accretion of satellites on
high angular momentum orbits could provide old stars in the thin disk,
and perhaps explain the Monoceros Stream.  However, the thin disk is
clearly far from a smooth, equilibrium system which could provide a
well-understood background population against which to define
substructure. Similarly to the situation in external galaxies, spiral
arms in the thin disk can cause significant disturbances in stellar
kinematics and positions that persist long after the perturbation has
passed (e.g.~de Simone et al.~2004), and resonant scattering can lead
to significant radial mixing (Sellwood 2008 and references therein;
Ro\v{s}kar et al.~2008). Resonances with the bar in the Milky Way can also
induce coherent motions (Dehnen \& Binney 1998).  These effects can
all give rise to  `moving groups' within which there is a large range
of stellar age and metallicity, consisting of stars that were not born
together, but have been perturbed to move together.  
Observational evidence has been provided by detailed analysis of space motions (e.g.~Famaey et al.~2005) and elemental abundances (Bensby et al.~2007b; Williams et al.~this volume). Indeed, it appears that all disk substructure, including the Monoceros Stream (Momany et al.~2006), can be produced by dynamical effects in the disk.  

\section{Cosmological Context}

The available stellar population evidence implies that the Milky Way
galaxy had a quiet history, with no significant mergers with dark
matter haloes since a redshift of $\sim 2$, where `significant' means
$\sim 20\%$ mass ratio to the disk, robust satellites, on non-circular 
orbits that take them into the region of the disk. This lack of merging is
apparently unusual in $\Lambda$CDM, where subsystems are typically
more concentrated than their hosts, and typically indeed on
radially biased orbits.  External disk galaxies also often have a thick disk component, and this component is again old (Mould 2005; Dalcanton, Seth \& Yoachim 2007). 

Gentle accretion must have dominated the mass
build-up of the Milky Way, with fluffy, gas-rich systems being the
primary means of matter infall. 
Late mergers are clearly contributing to the outer stellar halo, 
e.g.~the Sagittarius dwarf spheroidal.  Late accretion is perhaps building up  the gaseous disk, in the form of high velocity clouds. 
With accretion dominated by gas-rich systems, most star formation will
have occurred {\it in situ\/}, consistent with the high metallicity of
the bulge, and of the thick and thin disks, the components that dominate the
stellar mass of the Milky Way.   The first detections of CO emission lines in massive disk galaxies at redshifts $z \sim 1.5$  also imply {\it in situ\/} star formation (Daddi et al.~2008), as inferred for the Milky Way (and thus satisfying the Copernican Principle).

However, we do still lack important knowledge of the large-scale stellar populations of the Galaxy, including the  thick and thin disks far from the Sun.  Our knowledge of how the different stellar components are connected is also far from satisfactory.  There is a continuing need for 
large-scale spectroscopic studies, 
at both medium resolution, for broad kinematics and metallicities across the Galaxy, and high-resolution, for detailed elemental abundances and tracing streams.  There is also a need for comprehensive surveys of M31 and M33, to place the results for the Milky Way in a better context. 
The proposed Gemini/Subaru MOS instrument WFMOS will play an important part in this endeavour.

Bengt Str\"omgren emphasised the role played by the stellar populations of the Milky Way Galaxy in guiding our models of galaxy formation. That this remains true is testament to his legacy.

\acknowledgements {I thank the organisers for financial support, the Aspen Center for Physics for a stimulating environment, Greg Ruchti for help with Fig.~1 and him and other  collaborators for allowing me to show results in preparation.}

\end{document}